# Identifying the underlying structure and dynamic interactions in a voting network

Serguei Saavedra[a], Janet Efstathiou[a], Felix Reed-Tsochas[b]

[a]Department of Engineering Science, Oxford University, Oxford OX1 3PJ, UK
[b]Said Business School, Oxford University, Oxford OX1 1HP, UK

**Abstract**

We analyse the structure and behaviour of a specific voting network using a dynamic structure-based methodology which draws on Q-Analysis and social network theory. Our empirical focus is on the Eurovision Song Contest over a period of 20 years. For a multicultural contest of this kind, one of the key questions is how the quality of a song is judged and how voting groups emerge. We investigate structures that may identify the winner based purely on the topology of the network. This provides a basic framework to identify what the characteristics associated with becoming a winner are, and may help to establish a homogenous criterion for subjective measures such as quality. Further, we measure the importance of voting cliques, and present a dynamic model based on a changing multidimensional measure of connectivity in order to reveal the formation of emerging community structure within the contest. Finally, we study the dynamic behaviour exhibited by the network in order to understand the clustering of voting preferences and the relationship between local and global properties.

*Keywords: Structure; Network; Q-Analysis; Dynamics; Community Structure*

## I. Introduction

This paper investigates the static and dynamic relations within a complex voting network by analysing changes over time in both the network structure and certain multidimensional properties. We believe that many of the characteristics exhibited by the network can be identified by studying the structures formed by the flow of votes through the system. Here the structures should be understood as the various topological configurations or "backcloth" that result when threshold values applied to the weighted network are changed. For this approach we will combine network analysis techniques with Q-Analysis [1-2].

This research is based on the Eurovision Song Contest, since it allows us to study the complex structure of social interactions generated by the behaviour underlying a competition scheme on the basis of accessible empirical data. We analyse a 20 year window from 1984 to 2003. We have chosen to study this window because the explicit rules governing the contest are fixed during that time. The data set for the Eurovision contest is available on the web, and records the points exchanged between pairs of countries in each contest. One of the most attractive characteristics of the Eurovision data set is that it reflects an underlying complex social interaction, where the structure and dynamics describe the relations emerging from the exchange of points between countries. With some care this may be extrapolated to other cases in which a key mechanism is the exchange of material, information or goods.

Various researchers have studied the formation of cliques within the Eurovision voting community [3-5] and tested how reliably expert evaluation of musical quality are reproduced by this voting mechanism [6]. However, just a few of these studies have analysed the evolution of the network and how the links change over time [5]. In this paper we study different possible network configurations emerging from the multidimensional connectivity of the system. From this dynamic analysis, we identify what structure and dynamic changes can describe the emerging community structure and the relationship between the local and global properties of the network.

An interesting study concerning dynamic effects and behaviour in the Eurovision contest is that of Fenn et. al. [5], which is based on the voting patterns via a network analysis in order to describe some local and global properties within this system. Fenn et. al. found that this kind of network presents a particular clustering effect which leads to the formation of cliques between countries, and therefore the exchange of points is characterised by considerable biases and does not represent a fair contest. However, as acknowledge by the authors this result may be somewhat unrealistic since it is based on the assumption that all of the songs have the same quality, and therefore the same possibility of winning the contest.

We recognise that quality may be an unambiguous measure, since countries sharing geographical and cultural backgrounds may tend to unify musical concepts. In this paper we simplify social influence by focusing our analysis on finding the characteristics associated with becoming a winner based on the structure, relations and the global behaviour exhibited by the network. However, we must acknowledge that our approach looks at the behaviour and structure of this system assuming that it is only based on the pattern of social interactions, and does not take into account the very real prospect of tactical voting.

In the following section, we give more details about the mechanism of the Eurovision Song contest. In sections 3 and 4 we describe some basic concepts of social network analysis and Q-Analysis respectively. Section 5 is devoted to identifying a common structure shared by the winners throughout the years. Section 6 addresses the formation of cliques and their link to network topology. In section 7, we analyse the dynamic behaviour of the winner based on some principles of Q-Analysis. Finally, in section 8 we give our general conclusions.

**II. Empirical model: Eurovision Song Contest.**

The Eurovision Song Contest is an annual European event that in 2005 celebrated its $50^{th}$ anniversary. Currently, more than 35 countries participate and the contest has diversified to also include non-European nations. The rules, countries, voting scheme, and language of this contest have evolved over time. Nevertheless, from 1975 to 2003 a common format has prevailed, and although the number of countries varied between 18 and 26, the same voting scheme was preserved throughout these years.

For the purposes of this study, we use a 20 year window from 1984 to 2003 during which the voting rules remained constant. Throughout these years the voting rules of the contest remained unchanged, where each participating country performs a song and then is awarded points by other participating countries. Each country awards points to ten others on the scale {1,2,3,4,5,6,7,8,10,12}. The judges or countries award 12 points to their favourite song, then 10 to their second, down to 1 point. At the end of the contest, the country with the highest total number of points wins. This simple but effective voting scheme can be represented and studied through statistical and structural techniques in order to better understand the structure and dynamics of the network. In addition, this analysis shows the various conditions that a particular country needs in order to win this kind of contest.

**III. Network Analysis basics**

A graph or a network *G* (for the purposes of this work we are going to use the terms interchangeably) is an ordered pair of disjoint sets (*V*,*E*), such that for every graph *G*(*V*,*E*) there exists a set *V* of vertices and a set *E* of edges [7-9]. The relationships within a network can be represented either by edges (lines with no direction) giving rise to undirected graphs or by arcs (lines with a particular direction represented by an arrow) producing directed graphs. In addition, we may assign values or weights to the lines to give weighted or unweighted graphs.

Networks can be characterised by their local and global properties. The former focuses on individual properties of vertices compared to the rest of the graph, which in network sociology is frequently referred to as ego-network analysis. The latter is focused on the analysis of the entire network's structure and behaviour. Nevertheless, an analysis that takes into account both properties is undoubtedly necessary for a complete understanding of the network's dynamics and behaviour. Most of the measures used to understand network properties come from social sciences [10-12] and have been generalised by others [13-16]. Furthermore, if we aim to analyse the dynamics of the system, alternative approaches exist based on the growth and evolution of networks [15,17-18]. Many ways exist to represent a network; therefore, we must be aware of the particular characteristics of the system and what kind of information we are looking for. For instance, taking into account the flow and direction of links, a graphical representation of the Eurovision Song Contest network looks like Figure 1.

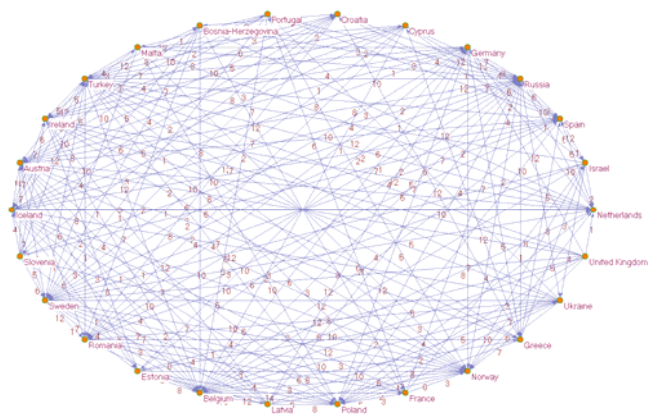

Figure 1. Circular layout of the 2003 voting network. Representation of a weighted and directed graph. Numbers on lines represent the total number of points exchanged by two countries.

## IV. Q-Analysis basics

In this paper we have decided to augment standard forms of network analysis with Q-Analysis because the latter allows us to analyse the properties of networks at different levels of connectivity. Q-Analysis takes a different perspective from traditional network theory, since it distinguishes a set of multiple geometrical structures defined by the number and type of interactions within the network.

Q-Analysis was developed by Atkin [1] and is an algebraic technique used in the first instance as a form to describe social structures. This technique is applied to reveal the relationships between two different sets $A = \{a_1, a_2, a_3, ...a_n\}$ and $B = \{b_1, b_2, b_3, ...,b_n\}$, where for every $a_i \in A$ and $b_j \in B$ there exists a $\mu$ relation between these two if they satisfy a particular relation rule. Therefore, we obtain the solution set of every relation $A \times B$ related to $\mu$. This can be represented by an incidence matrix $M$ showing the interactions between the sets:

$$M = \begin{array}{c} \\ a_1 \\ a_2 \\ a_3 \\ . \\ a_n \end{array} \begin{array}{cccc} b_1 & b_2 & b_3 & b_n \\ \left| \begin{array}{cccc} 1 & 0 & 1 \ldots & 1 \\ 0 & 0 & 1 \ldots & 0 \\ 1 & 1 & 1 \ldots & 1 \\ & & & \\ 0 & 1 & 0 \ldots & 1 \end{array} \right| \end{array}$$

where the $\mu$ relationships are represented by the elements $r_{ij}$ of the binary matrix defined by
$r_{ij} = $ 1 if $a_i$ is $\mu$-related to $b_j$,
     0 otherwise

The most important concept behind Q-Analysis is the possibility to define the structures or relations of the system in terms of a geometrical representation [19-22]. This geometrical representation is a unique complex which takes into account the multidimensional characteristics of the network. This complex is called a simplicial complex [1] and it is defined as $K_A(B, \mu)$, where the elements of the set $A$ are the simplices $\sigma_p(a_i)$ and the elements $B = \{b_1, b_2, b_3, ..., b_n\}$ of the simplices are the vertices. For instance, for a complex $K_A(B, \mu)$ with simplex $\sigma_{ai}\{b_1,b_2,b_3,b_4\}$, the dimension of the element $a_i \in A$ which is $\mu$-related to $b_1$, $b_2$, $b_3$ and $b_4$ is 3 (one less than the number of vertices). This simplex may be represented by a tetrahedron. The group of simplices forms the simplicial complex. This gives the means to study complex systems in their natural multidimensional structure, which provides new insights since we can analyse separately their properties at different dimensions [22].

Related to the idea of multidimensional structures, each simplex from $K_A(B, \mu)$ is related to a number of elements of the set $B$, and is $q$-near to other elements within its group according to the shared vertices. Hence, $q$-connected chains can be formed through common geometrical faces such that for a finite sequence of simplices $\{\sigma(a_1), ..., \sigma(a_n)\}$, the following is satisfied:

1. $\sigma(p)$ is a face of $\sigma(a_1)$,
2. $\sigma(a_n)$ is a face of $\sigma(r)$,
3. $\sigma(a_i)$ and $\sigma(a_{i+1})$ have a common face of dimension $\beta_i$, for i= 1,…,n and
4. $q=\min \{p, \beta_1, ..., \beta_n, r\}$.

Therefore, *q*-near refers to two simplices sharing at least *q*+1 vertices; and two simplices will form a *q*-connected chain if there is any pairwise *q*-near simplex between them. For example, if we have the following simplices σ($a_1$)= <$b_1,b_2$> σ($a_2$)=<$b_1,b_2,b_3$> and σ($a_3$)= <$b_2,b_3$>, they form the complex shown in Figure 2. From this figure we can appreciate that at dimension 1 (lines) they form a weak link since $a_3$ and $a_1$ are not related directly and just through $a_2$. While at dimension 0 (points) all of them share the vertex $b_2$ and therefore we can say that they form a strong link. Hence, according to the previous definition of *q*-connectivity, these simplices will be related in each dimension as summarised in Table 1, where at all dimensions there is a single class or chain but with different strengths or contents. This relational and multidimensional concept has given Q-Analysis its uniqueness and strength.

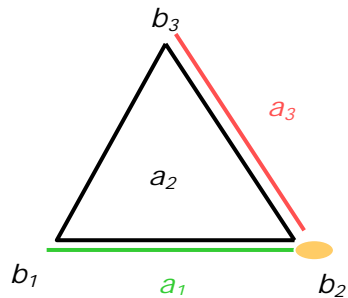

Figure 2. Graphical representation of the relationships at the simplicial complex $K_A(B, \mu)$. Here, $a_1$, $a_2$ and $a_3$ are the simplices and $b_1$, $b_2$ and $b_3$ are the vertices. This representation shows that $a_2$ is a triangle whereas $a_1$ and $a_3$ are lines. It is interesting to note that the only vertex they share in common is $b_2$ which makes them strongly connected at dimension 0.

| Dimension | Classes | Type of relation |
| --- | --- | --- |
| 2 | {$a_2$} | Strong |
| 1 | {$a_1,a_2,a_3$} | Weak |
| 0 | {$a_1,a_2,a_3$} | Strong |

Table 1. Summary of the q-connectivity structure shown in Figure 2. A relation is strong when all the element is the class are related directly, whereas in a weak relation an element acts as a link between two other elements.

The summary of the relations in a system is defined by the *Q*-structure vector, where for a given complex K, each dimension has a corresponding number of classes $Q_q$. This vector is used to gain some insights on the connectivity and dimensions of the network structure [2].

$Q = \{ Q_{(dim\ K)}, Q_{(dimK-1)},\ldots,Q_{(0)}\}$

For our previous example, the *Q*-structure vector is:

$$Q = \{\overset{2}{1}\ \overset{1}{1}\ \overset{0}{1}\}$$

where the dimension is indicated above the number of classes. As we can see, each of the dimensions is formed by a single class as shown in Table 1.

In addition, another vector taken from Q-Analysis is the $Q^*$-obstruction vector, which defines the level of communication or interaction among vertices in each dimension. The obstruction vector quantifies the number of obstacles or gaps found in a particular dimension [23]. High values indicate that it is a fragmented or disconnected structure. This vector is obtained by subtracting a unit vector $U$ from the $Q$-structure vector. Calculating the $Q^*$ vector for our example, we obtain

$$Q^* = \{ \overset{2}{0}\ \overset{1}{0}\ \overset{0}{0} \}$$

where the dimension is indicated above the number of obstructions. In this case, the level of obstruction is null and from this result we can draw two different conclusions. On one hand, we can say that this is a stable system with no opportunity to changes (e.g. authoritarian regime); while on the other hand, we can argue that this is the perfect medium for communication where no obstruction exists at all (e.g. democratic regime) [1].

Q-Analysis allows us to study two different dynamic approaches; one that is centred on classical dynamics where the system is subjected to externally generated perturbations and another that is focused on modern dynamics where the entity that is changing is the system itself. These approaches have led to better dynamic studies [20], and help to understand the backcloth and traffic of the system or as Duckstein [23] describe them: *"the relatively static and relatively dynamic aspects of the system"*. It is easier to see the backcloth as a particular defined set of agents and their relations, and traffic as the flow and outcome of the system. Therefore, the backcloth is understood as the actual structure of the complex, while the traffic is the relation of patterns or forces acting on the backcloth through different periods of time. The application of these properties has been widely spread over many disciplines in which the aim is to describe the behaviour of particular systems and establish possible outcomes [21].

We have briefly described the basic concepts of the techniques used in this paper. The following sections will move on analysing the data answering the main concerns under study.

## V. The role of structure

Is there any way to identify the real quality of a song? Which are the main characteristics that make a country become a winner? These are some of the basic questions people would like to ask about this kind of contest. We address these questions by a structural methodology that focuses on finding a common structure which identifies the winner based on the global topology of the network over the time, and that generates useful information about some of the main characteristics of a winner.

The dynamic properties of the network are the most important characteristics to account for the real behaviour of the network. We present a methodology based on Q-analysis in order to find the winning structure, and compare it to static techniques. This section starts by defining what we mean by quality. Two static techniques are used as benchmarks; one is based on spatial representation and the second one on the degree distribution. Finally we present the results of a dynamic analysis centred on Q-Analysis.

A common definition of quality is that of the capacity of a country to attract votes, usually measured by the total number of points (weighted graph). Of course, we know that quality is a rather subjective measure, yet we believe that a fair approach to make it a global property can be achieved by counting the number of participants who vote for a particular song regardless of the points they have awarded (unweighted graph). Therefore, we define a vote as the binary relationship between two countries which generates an unweighted network.

We assume that quality is not only revealed by achieving the highest score; but also, in a fair contest, this quality should be recognised by the support of the majority of the participants, i.e. almost every country should vote for the best song regardless of the points. In order to test this idea, we examine the graphical representation. Here we analyse an unweighted graph, and therefore the winner should be located in the middle of the network because of its high number of votes. This has been performed using the Fruchterman-Reingold [24] layout algorithm whose function places the most connected nodes in the middle of the graph (Figure 3).

From the graphs below, we can see that in 2003 and 2001 the winners, Turkey and Estonia, present a different topology which makes very hard to tell who the winner is in each year. This makes necessary to use different network metrics in order to tackle our questions: Is the winner supported by the number of votes? How does this affect the definition of quality?

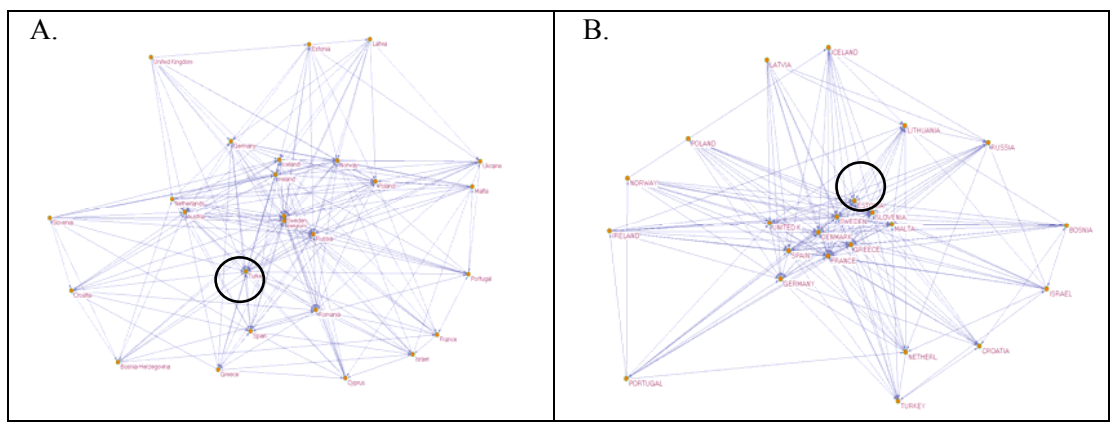

Figure 3. Fruchterman-Reingold [24] layouts. The algorithm places the most connected nodes in the middle of the graph. A and B stand for the 2003 and 2001 data respectively. Note that the winners, inside the circles, present different topologies and are not located in the middle of the network. The visualisation software used throughout this study is Pajek [25].

We analysed the distribution degree in order to determine the correlation between the degree (quality) and the winner. As we have stated before, we are interested in establishing to what extent is the winner supported by the other participants; therefore, we have studied the directed in-degree distribution. The in-degree of a vertex is the number of directed links towards itself, i.e. votes achieved. We would expect to find that the winner has the highest degree rank, and therefore we expect the winner to obtain the highest correlation value among all the countries.

We have developed a correlation analysis for each year from 1984 to 2003 using the Spearman's rank correlation coefficient, where we measured the relation of the overall position, i.e. the winner with its corresponding quality rank. This coefficient is suitable for measuring the strength between two variables taking into account their ranks, and is defined by:

$$R^2 = 1 - \frac{6\sum d^2}{n^3 - n}$$

where $R$ is the Spearman's rank correlation coefficient, $d$ is the rank difference between two variables and $n$ is the sample size; with $R \in [-1,1]$ and the closer $R$ is to +1 or -1, the stronger the correlation.

For the winner we have a correlation value of 0.586 with a 95% confidence level. This is a positive correlation between the winner and the number of votes, but cannot be considered a guaranteed to identify the winner. We have proved that the unweighted network is not sufficient to identify the winner in a dynamic environment.

In order to improve our understanding of the properties which define a winner, we focus on the backcloth and traffic of the system [1]. Following this idea, we defined the backcloth as the relatively stable structure of the system, and traffic as the particular outcome achieved. Since these data are centred on the votes among countries and we already know that the winner is not always the one with most votes (recall the in-degree distribution), it is important to determine what kind of voting structure makes a winner different from the rest of the participants. Although Q-Analysis defines the structure of two different sets, it is possible to work with a single set since the voting table can be seen as two different sets which are not symmetric, i.e. the rows are the countries receiving votes, and the columns are the countries giving votes.

The variation of the relational structure is achieved by changing the $\mu$-relation rule by means of different threshold parameters, best known as "slicing" the matrix. The slicing procedure consists in declaring a binary relation between countries A and B if country A awarded points to country B more than or equal to the threshold value. We hold the definition of quality as the number of votes; yet, while changing the threshold we are just modifying the concept of votes. The winner is identified within the $Q$-structure vector and the corresponding threshold is classified according to the winner's dimension or position i.e. if the winner has the highest dimension or number of votes within the $Q$-structure, then this particular slicing threshold will be classified as one (1=winner). This approach is based on Central Place Theory [26] where the central place (winner) is identified by its relevant position or dimension degree, and the varying parameters are countries and votes.

We performed a Q-Analysis for the complete data varying the slicing threshold from 1 to 12, and we find that the unique winner structure is achieved with a threshold of six. We again use again Spearman's rank correlation to measure the reliability of a threshold to recognise the winner over the years. From Figure 4 we can see how the different structures struggle to identify the winner while threshold six is the only one that recognises the winner with a perfect correlation value. Therefore, we can claim that a standard definition of quality, or at least something that can show us the real quality of a song in such a multicultural event, is the number of votes awarding six points or more. Furthermore, we have found a reliable structure capable of identifying the winner based just on the topology of the network. In addition, comparing the number of votes obtained by the winner through the *Q*-structure vector with threshold six to the maximum possible on the network (one less the total number of participants), we obtain on average a proportion of 0.621. This suggests that what truly characterises a winner is to be awarded within the five highest scores by just 60% of the participants.

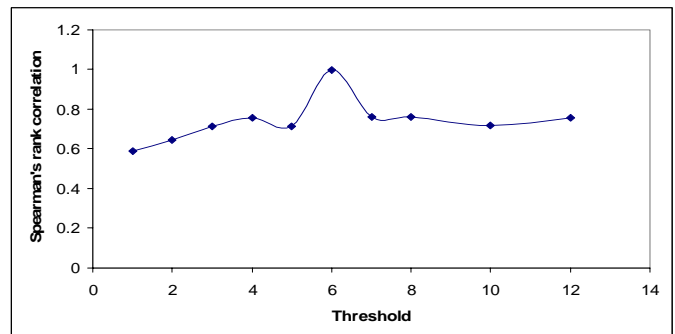

Figure 4. Slicing procedure used to detect the winning structure. The x-axis corresponds to each voting threshold, and the y-axis refers to the Spearman's correlation value for each threshold calculated over the entire study period. Note that threshold six is the only value that achieves to recognise the winner in all years (y-axis value=1).

Taking into account the dynamic behaviour of networks, we have identified a unique structure capable of identifying the winner of a contest throughout the years, while static techniques have struggled to achieve this task. Further analysis shows that the real quality of a song can be defined by the number of votes equal or above six, and what truly characterises a winner is to be awarded within the five highest scores by just 60% of the participants.

## VI. Community structure

The formation of cooperative groups is a well recognised property of complex socio-economic systems. However, the analysis of the structure and dynamics of such communities allow us to better understand the evolution of these systems [14]. Fenn at. al. [5] have demonstrated that the Eurovision contest is far from being a randomly generated network and that there are indeed some voting cliques defining the structure of the network. Here, we give further evidence that voting cliques are an evident characteristic of this voting network based on structural characteristics. We present a dynamic clustering procedure centred on a multidimensional structure in order to identify voting groups, and we test the reliability of this procedure based on a random network.

One of the most widely used metrics in social network analysis is the clustering coefficient, which reflects the increased probability that two individuals that share a common friend have a greater than random probability of being friends themselves [14]. Following the idea of triad structures [12] that recognise cliques within groups, we have analysed structures that can give strong evidence of the formation of groups. The analysis of local structures has been generalised under the name of network motifs [32]. We quantified the presence of three kinds of triad throughout the years and compared their results to those of a random network.

The first triad (Figure 5-a) called 300 in the M-A-N labelling system [27] refers to that structure with reciprocal links (A votes for B and B votes for A) among three different countries. This structure not only reveals a function similar to the clustering coefficient but also gives stronger evidence of union, since the probability of finding such a structure should be very low. The second triad (figure 5-b), called 102, refers to the relationship between two countries sharing a reciprocal link with no links to a particular third country. The third triad (figure 5-c), called 003, shows the opposite function as the 300 since within this triad no link exits between any country, and therefore gives evidence of a group division process taking place.

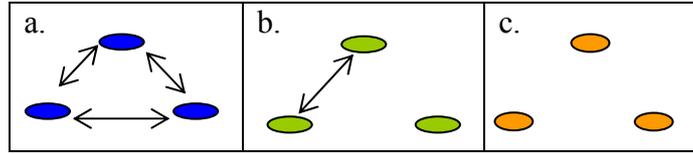

Figure 5. Triad structures, a is the perfect triad structure (300), b is the perfect dyad structure (102), and c shows clusterability (003).

The presence of triads can be established by calculating the respective frequency in a random network with similar number of links and comparing them to the number of triads found in the original network. Since we are taking into account triads or *3-subgraphs*, first we must establish the number of different possible triads and assign to each class of triad a probability of occurrence. We define $E(S_u)$ as the expected frequency of a *k*-subgraph in a triad class *u*. If we define $p_k(u) = $ (K in class *u*) as the probability of a *k*-subgraph of being in class *u*, and since there are $\binom{g}{k}$ different triads in a network of *g* nodes, the probability of a class *u* is given by $\bar{p}_u = \dfrac{1}{\binom{g}{k}} \sum_K p_k(u)$ and $E(S_u) = \binom{g}{k} \bar{p}_u$.

We have calculated the frequency of a triad based on its given probability. According to the definition of a random graph, each outcome is completely independent of any other. Since in the Eurovision Song Contest each node has a fixed out-degree of ten, the probability of a link from A to B is given by $P_{A \to B} = 10/N$, where there are *N+1* nodes in the network. Hence, the probability of any *3*-subgraph being in class 300 (Figure 5-A) is given by

$$P_3(300)=[(P_{A \to B}) \cap (P_{A \to C})] \cap [(P_{B \to A}) \cap (P_{B \to C})] \cap [(P_{C \to A}) \cap (P_{C \to B})] = \left(\frac{100}{N^2}\right)^3,$$

and the probabilities of being in class 120 and 003 respectively are:

$$P_3(120)=\left(\frac{100}{N^2}\right)\left(1-\frac{10}{N}\right)^4 \quad P_3(003)=\left(\frac{N-10}{N}\right)^6$$

In order to be able to compare different years and triads, we calculated the relative difference of each structure defined by [12]:

$$ro = \frac{oc - ex}{ex}$$

where *ro* is the relative occurrence, *oc* is the actual occurrence (calculated with Pajek [25]) and *ex* is the expected frequency in a random graph, in such a way that non zero values indicate deviations from randomness and high positive values represent the dominance of such triad.

Table 2 shows the relative difference for the three triads in each year. From this table, we can appreciate that the number of observed triads 300 and 003 is significantly higher than random. This gives more evidence of a group division process and the formation of cliques. However, the triad 120 shows a frequency below random, suggesting that the formation of cooperative voting groups is to be found mainly among three or more countries rather than between two.

In addition, we looked at the dynamic behaviour of voting groups analysing the perfect clustering structure (300). We observe that the relative frequency of this triad grows and changes throughout the years (Figure 6). It may be that for some countries the songs are becoming increasingly similar, that is reinforcing the propensity of these countries to vote for each other. Also there is the possibility that these triads are formed as a consequence of the various divisions (e.g. political, geographical) taking place in the wider world beyond the Eurovision Song Contest. Further, this growth can be explained by the constant addition of new countries which identify themselves with a particular group. This is the case in 1994, when a new group of countries entered the contest (Estonia, Hungary, Lithuania, Poland, Romania, Russia, Slovakia). As we can see from Figure 6, this has had an impact and changes the configuration of these networks.

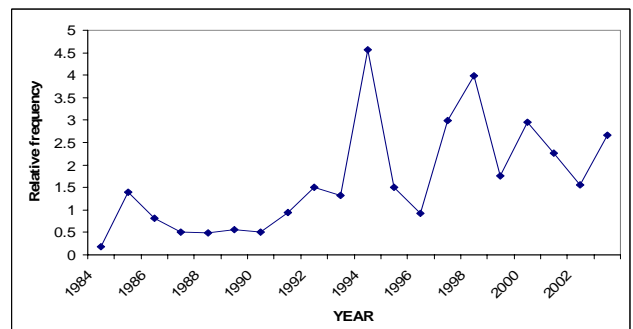

Figure 6. Graphical representation of the relative frequency of the balanced triad 300 from 1984 to 2003. Note the peak in 1994 when a new group of countries entered the contest (Estonia, Hungary, Lithuania, Poland, Romania, Russia, Slovakia).

| Year | '84 | '85 | '86 | '87 | '88 | '89 | '90 | '91 | '92 | '93 | '94 | '95 | '96 | '97 | '98 | '99 | '00 | '01 | '02 | '03 | Avg. |
|---|---|---|---|---|---|---|---|---|---|---|---|---|---|---|---|---|---|---|---|---|---|
| Triad 102 | -0.4 | .57 | -0.1 | -.37 | -.57 | -.04 | -.16 | -.15 | .21 | -.44 | -.11 | .06 | .27 | -.07 | -.31 | -.38 | .08 | -.07 | .64 | .03 | -0.06 |
| Triad 300 | .19 | 1.3 | .82 | 0.5 | .49 | 0.5 | 0.5 | 0.9 | 1.5 | 1.3 | 4.5 | 1.5 | .92 | 2.9 | 3.9 | 1.7 | 2.9 | 2.2 | 1.5 | 2.6 | 1.66 |
| Triad 003 | 1.5 | 1.1 | .94 | .67 | 1.4 | .26 | .76 | .92 | .91 | .63 | .96 | 1.0 | .44 | .94 | .92 | .93 | 1.2 | 1.9 | 1.1 | 0.9 | 0.98 |

Table 2. Relative frequencies from 1984 to 2003 of the three triad models. The highest average value is found at the 300-triad (last column).

The first part of this section analyses voting groups based on a purely structural configuration giving strong evidence of the formation of cliques. In order to define a community structure, we propose to take into account the information carried by the links. Therefore, we have studied the formation of clusters or voting groups according to their dynamic structures.

Macgill [28] suggests Q-Analysis as a technique that can be used to find the different groups which are formed on a network. However, his approach is focused just on a static structure. By contrast with the static approach, we have developed a cluster analysis based on Q-Analysis taking into account the derived structures drawn by a distinct weight criterion. The study has been centred on the data from 1994 to 2003 where the formation of clusters appears to be quite strong and seems quite erratic (Figure 6).

The analysis has been based on forming groups according to their connectivity and dimension while changing the threshold parameter from 12 to 1 by slicing the matrix. Hence, a link between A and B is generated if A awards points to B which are equal or greater than the threshold value. Countries are better connected if they share a class in a high threshold and high dimension (similar countries voting for them). This procedure generates clusters as countries start to appear linked in $q$-chains. The clustering procedure starts with the highest threshold value and then decreases the value until every country belongs to a particular cluster. The matrix used for this case has been obtained from the sum of points that each country gave to each other divided by the number of years they competed. Also, the diagonal values have been changed from 0 to 14. This has been done in order to account for the real connectivity, i.e. every country starts with 0-dimension, and if two countries voted for each other then at 1-dimension they are in the same class; otherwise, no connection exists. Therefore, we have just been interested in classes appearing above the 0-dimension. The Q-Analysis has been performed for the complex $K_A(B, \mu^{-1})$ which is the inverse relation of the original matrix and takes into account the votes received by each country. This approach is based on classic dynamics where an external factor (threshold) is changed and the output results are quantified.

Figure 7 shows the result of the clustering analysis where ten groups are formed. The information is presented as a dendrogram and shows the formation of clusters according to their appearance while varying the dimension and threshold. It is interesting to note that all the countries are clustered together after reaching threshold six, which supports the idea of a structure capable of linking all the countries into a single group as a homogenous criterion. Also, we can observe that Latvia and Estonia as well as Cyprus and Greece are the best connected pairs in the network. Surprisingly, France is quite different from the rest of Europe, in line with the results of Fenn et.al. [5]. Another interesting fact is that almost all the countries from a particular group belong to the same geographical area, which presumably reflects similar preferences of culture and taste exist within these countries, or some kind of political or other relations. Further discussion of the nature of the cliques is left to the cautious reader.

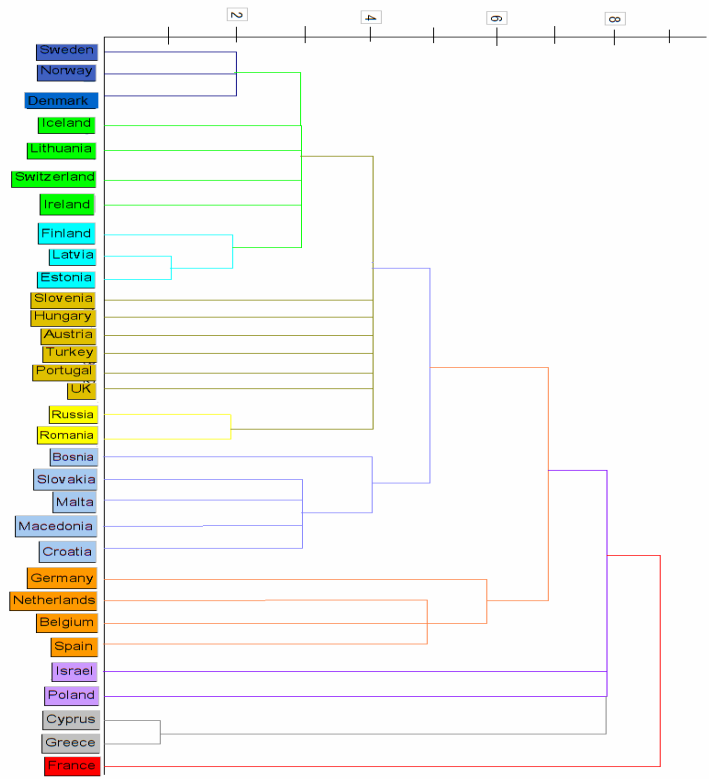

Figure 7. Dendrogram of the voting clusters from 1994 to 2003. Each colour represents a group or clique. We gave a value of 1 for the highest connectivity found among countries, and this is incremented while lowering the connectivity criterion in order to link new countries.

Further, we have investigated if this number of cliques is an accurate division for this network, and whether another number of cliques might have better represented the network's structure. To test our division, we used Newman's modularity [29] which is a measure of the quality of a particular division based on the expected proportion of edges within a community in a random network compared to the real one. This modularity $Mo$ is defined for a weighted graph as:

$$Mo = \frac{1}{2m} \sum_{ij} [A_{ij} - \frac{k_i k_j}{2m}] \delta(c_i, c_j)$$

where $m$ is the number of edges, $A_{ij}$ are the points given from country $i$ to country $j$, $k$ is the degree of the vertex and $\delta(c_i, c_j)$ equals 1 if $c_i$ and $c_j$ belong to the same community, 0 otherwise.

A value of 0 indicates that the network is not better than random and the maximum value for $Mo$ is 1. Newman suggests that non zero values of $Mo$ indicate variations from randomness and values between 0.3 and 0.7 are indicators of good divisions.

The division of communities has been taken from the previous dendrogram (Figure 7) where at first we start with a single community and after slicing the connections, we end up with ten different communities.

After performing the analysis (Figure 8) we observe that a single community is not feasible; and values above 0.3 are achieved from a division of four groups, while the highest value of *Mo* (0.459) is achieved for a division of eight groups. This confirms the accuracy of our previous division, retaining the groups of {Sweden, Norway, Denmark}, {Iceland, Lithuania, Switzerland, Ireland}, and {Finland, Latvia, Estonia} as a single cluster. A final interesting observation is that this last group contains not just countries sharing similar geographical areas, but also the 70% of the winners within this 10 year period.

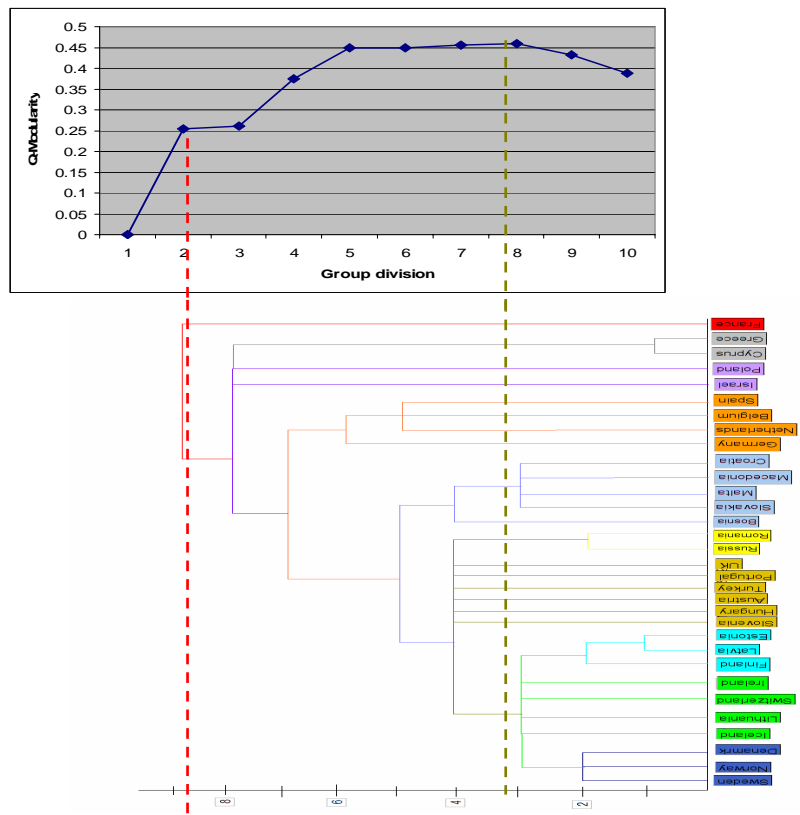

Figure 8. Relationship between Newman's modularity values and the dendrogram obtained for the community structure. Note that the maximum modularity value is achieved with a division of eight groups (brown dashed line).

## VII. Dynamic behaviour

We have already established that a relatively static structure defined by its own global configuration can be applied to adequately find the winner of any contest within this period of time. In this section we study the local structural properties of the winner and how they are related to the global capacity of the network to establish a homogenous quality criterion.

This study focuses on the relation among countries and the strength or dominance of the winner within the structure. First, we observe the different structures formed in each year from 1984 to 2003 in order to gain some insights on how the winner is related to the other participants. For this approach we analyse for all years the *Q*-structure vectors with a threshold of six.

Table 3 shows two samples of structure and obstruction vectors with their elements and relations. It reveals two interesting facts; one is the dominance of the winner, and the other is how the network generates a division of preferences expressed by the different classes in each dimension. In addition, we can see that the higher dimensions behave in a completely different manner. For example, in 1996, Ireland was the only country in the last five dimensions. In 1988, Switzerland, the winner, stands by itself only in the last dimension. We propose to analyse this kind of behaviour using the *Q\**-obstruction vector which defines the diversity or obstruction to flow within a network. From the same table, we can see that although in 1996 the dimension of the *Q*-structure is bigger than in 1988 (15 to 12), the obstruction to change is slightly higher in 1988 due to a higher dimension and number of classes at the *Q\**-obstruction vector. This is twofold; first, the difference in 1996 between dimensions of the *Q*-structure and *Q\**-obstruction vectors tells us that the last five dimensions are entirely dominated by Ireland. Second, the higher obstruction values of 1988 show a higher diversity of preferences.

|  | Year 1996 | Year 1988 |
|---|---|---|
| **First five places (1-5)** | Ireland, Norway, Sweden, Croatia, and Estonia | Switzerland, UK, Denmark, Luxembourg, and Norway |
| ***Q*-structure** | 15  0<br>*Q*(1 1 1 1 1 2 3 3 4 5 7 7 2 2 2 1) | 12  0<br>*Q*(1 2 2 2 2 6 8 8 4 4 2 1 1) |
| ***Q\**-obstruction** | 10  0<br>*Q\**(1 2 2 3 4 6 6 1 1 1 0) | 11  0<br>*Q\**(1 1 1 1 5 7 7 3 3 1 0 0) |
| **1 – connectivity** | {Iceland, Sweden, UK, Turkey, Netherlands, Ireland, Greece, Norway, Portugal, Cyprus, Malta, Croatia, Austria, Estonia, Poland}<br>{ Slovenia} | {Sweden, UK, Turkey, Spain, Netherlands, Israel, Switzerland, Ireland, Germany, Denmark, Norway, Luxembourg, Italy, France, Yugoslavia} |
| **6 – connectivity** | {Sweden, Ireland, Norway}<br>{Portugal}, {UK}, {Croatia}, {Estonia} | {Switzerland, Norway}<br>{Israel}, {Denmark}, { UK}<br>{ Luxembourg }, { Netherlands },<br>{ Ireland }, { Yugoslavia} |
| **7 – connectivity** | {Sweden, Norway}<br>{Ireland}, {Croatia}, {Estonia} | {Switzerland},{Norway}, {Denmark}, { UK}<br>{ Netherlands }, { Yugoslavia} |
| **8 – connectivity** | {Sweden}, {Ireland}, {Norway} | {Switzerland},{ UK} |
| **9 – connectivity** | {Sweden}, {Ireland}, {Norway} | {Switzerland},{ UK} |
| **10 – connectivity** | {Ireland}, {Norway} | {Switzerland},{ UK} |
| **11 – connectivity** | {Ireland} | {Switzerland},{ UK} |
| **12 – connectivity** | {Ireland} | {Switzerland} |
| **13 – connectivity** | {Ireland} |  |
| **14 – connectivity** | {Ireland} |  |
| **15 – connectivity** | {Ireland} |  |

Table 3. *Q*-structure and *Q*-obstruction vectors for 1996 and 1988. Sample of the *q*-connectivity for both years. For each q-dimension, the table shows *q*-chains of simplices (countries) inside the brackets, and classes (group of countries) separated by different brackets.

In addition, we have analysed the winner's dominance or connectivity differences throughout the years based on the winner's level of integration with the other participants. Following Casti [30], eccentricity is a valid measure to account for connectivity since it gives us the integration of a simplex into the structure, or how different a simplex is from the rest of the participants. As we need to account for the complete mapping or global vision of the structure, we have taken Chin's eccentricity [31] which measures the level of integration for each simplex by:

$$ecc(\sigma) = \frac{2\sum_i q_i/\sigma_i}{q_{max}(q_{max}+1)}$$

where $q_i$ stands for all the dimension where $i$ appears, $\sigma_i$ is the number of simplices on the class, $q_{max}$ is the highest dimension of the complex. *ecc* takes on values in [0,1], where high values indicate higher eccentricities.

We have calculated the global eccentricity of the winner using Chin's measure and modified $q_{max}$ as the maximum possible dimension of the complex in order to normalise the measure and to be able to compare different years. Therefore, by measuring the winner's eccentricity throughout the years with a threshold of six, we analyse whether it is a stable structure or if there is any intrinsic process shaping its configuration. High values indicate that the winner is an outstanding simplex and has a stronger dominance over the other participants. This measure reveals the strength of the global structure as seen from the winner's perspective.

Further, we measure the stability of the quality criterion by comparing the local eccentricity of the winner to that of its closest rivals. We analyse the natural division of preferences or groups and compare them to the global eccentricity. In this way, we have looked at the *Q*-structure differentiating its components and connectivity chains. Stability is measured by calculating again Chin's eccentricity for the winner minus the second place's eccentricity. In this case $q_{max}$ is just the highest dimension of the complex since we are interested in obtaining the local difference between the first and second place. The main idea behind this property is to measure the extent to which the winner is supported by a homogenous criterion or whether there is a clear voting difference. Therefore, high positive values from this difference are understood as a high homogenous criterion, a zero value as an indefinable criterion, and negative values as an unstable heterogeneous criterion.

We calculate the values for the winner's global eccentricity and the difference of local eccentricities using the data from 1984 to 2003 with a slicing threshold of six. Figure 9 shows that in this kind of networks the dominance of the winner is proportional to the criterion's homogeneity, and this is supported by a Pearson correlation coefficient of 0.867. In addition, the figure reveals that when there is not a clear winner, the network not only strives to differentiate who the winner is, but also the entire dimension of the network shrinks. This suggests that when there is a division of opinions, the natural tendency is to form at least two main clusters.

In order to validate this correlation and to ensure that we are not analysing dependent measures, we have performed the same analysis for a random network, where we start with *N* nodes and the probability of a link existing between two nodes is defined by:

$$P = \frac{5}{N}$$

where *P* is the probability, 5 stands for the out-degree (5 highest votes) and *N* is the number of participating countries for each year.

A first analysis consists of finding the correlation between the local and global eccentricities for a random simulation of each year (Figure 10), and a second consists of finding the correlation taking the average values for 100 different simulations of each year. In the first case, the correlation is 0.222, while in the second we obtained a negative correlation of -0.206. The two are clearly below the original correlation; and therefore, this provides evidence that the relation is not a property of a random network with the same characteristics.

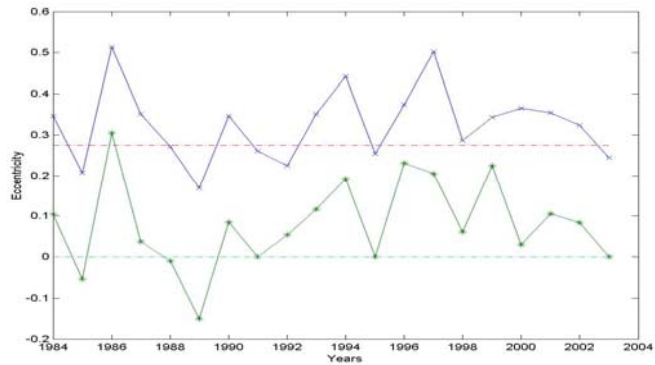

Figure 9. Relation between the winner's global eccentricity (blue solid line) and the difference of local eccentricities (green solid line). The red dotted line defines the global eccentricity threshold of 0.264 where the local eccentricities fall below zero. Note that both eccentricities present a positive proportional behaviour.

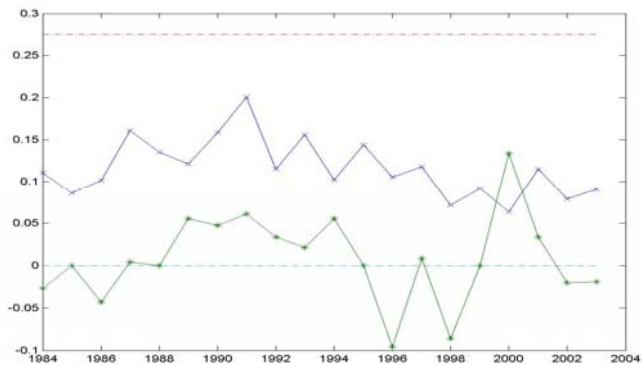

Figure 10. Random relation between the winner's global eccentricity (blue solid line) and the difference of local eccentricities (green solid line). Note that they do not present a proportional behaviour as the original values (Figure 9).

The second interesting fact emerges from this analysis when we look for those stability values equal or less than zero. We find that these values respond to a particular global eccentricity threshold of 0.264 (Figure 9), where below this threshold the stability of the quality criterion is zero or lower. We can illustrate this remark taking some examples from the contest itself:

a) 1985 (local eccentricity = 0.207, global eccentricity = -0.05): The winner achieved 52% of the maximum number of votes within the five highest points. The negative value can be explained by the high number of classes in the $Q$-structure vector.

$$\overset{10}{Q(2\ 3\ 4\ 5\ 7\ 5\ 2\ 1\ 2\ 2\ 1)}\overset{0}{}$$

b) 1991 (local eccentricity = 0.26, global eccentricity = 0): Sweden, the winner, achieved a good percentage of support since a 63% of the participants gave it votes within the five highest scores. Nevertheless the second place, France, achieved the same final score. In this year, there was a tie between these two countries that had to be decided using an extra criterion (number of votes with a 12 score).

$$Q(1\ 1\ 2\ 3\ 3\ 3\ 6\ 5\ 4\ 4\ 3\ 3\ 1\ 1)^{13}_{\phantom{13}0}$$

c) 1999 (local eccentricity = 0.34, global eccentricity = 0.22): The winner achieved a 69% of the maximum number of votes within the five highest points. The high positive value can be explained by the low number of classes in the Q-structure vector showing a uniform criterion.

$$Q(1\ 1\ 1\ 1\ 2\ 4\ 2\ 2\ 2\ 3\ 4\ 5\ 2\ 1\ 1)^{15}_{\phantom{15}0}$$

On the other hand, we can see some kind of exceptions to the proportional relation between these two properties; yet, it is interesting to note that these exceptions occur just in one direction, i.e. a high global eccentricity can appear with an unstable criterion but not vice versa. This was the case in 1987 and 2000 (Figure 9).

## VIII. Conclusions

We have found that the Eurovision Song Contest is characterised by complex local interactions and a well defined global pattern describing the position of the winner. This study has analysed these properties in order to understand the static and dynamic forces acting through these networks. We acknowledge the subjectivity of different measures (e.g. quality) and for this purpose we have proposed to look for a homogeneous structure that can give the same results throughout a study period. This is important since we needed to have a basis of comparison that can take into account the particular characteristics of each period within the same framework.

We have concluded that what truly characterises a winner in this voting network is to be awarded within the five highest scores by just 60% of the participants. Further, we have seen how the inclusion of new countries in the contest since 1994 has dramatically influenced the behaviour of the voting network while the formation of clusters has become more evident. Also, we have presented a dynamic clustering methodology which is based on the strength of the links of a particular structure and on the changes suffered due to dynamic behaviour over a period of time. We have given further evidence of clique formation, and found that clustering is a phenomenon that appears mainly between countries sharing a geographical area.

Finally, studying the dynamic behaviour of the network, we have applied a structural approach using Q-Analysis in order to describe this property. This analysis consists of measuring the relation between the dominance of the winner and the level of the homogenous criterion to choose this winner. We find that there is a proportional relationship between these two properties, and a particular threshold is needed in order to achieve a clear victory.

## Acknowledgments

We would like to acknowledge helpful discussions with Ani Calinescu and Neil Johnson as well as other members of the CABDyN Research Cluster at Oxford University. S.S. holds a research studentship funded by the European Commission under the NEST Pathfinder Programme via the MMCOMNET Project (Contract No. 012999).

## References

[1] R. Atkin, *Mathematical Structure in Human Affairs*, London: Heinemann Educational, 1974.
[2] J. Beaumont, A. Gatrell, *An Introduction to Q-analysis*, Norwich : Geo Abstracts, 1982.
[3] G. Yair, "Unite Unite Europe" The political and cultural structures of Europe as reflected in the Eurovision Song Contest, *Social Networks* **17** (1995) 147-161.
[4] D. Gatherer, Birth of a meme: The origin and evolution of collusive voting patterns in the Eurovision Song Contest, *Journal of Memetics – Evolutionary Models*, (2003), **8**.
[5] D. Fenn, O. Suleman, J. Efstathiou and Neil. Johnson. How does Europe Make Its Mind Up? Connections, cliques, and compatibility between countries in the Eurovision Song Contest, *Physica A* **360** (2006) 576-598.
[6] M. Haan, S. Dikkstra and P. Kijkstra, Expert Judgment Versus Public Opinion – Evidence from the Eurovision Song Contest, *Journal of Cultural Economics* **29**, (2005) 59-78.
[7] B. Bollobás, *Graph Theory : An Introductory Course*, New York : Springer-Verlag 1979.
[8] R. Wilson, *Introduction to Graph Theory*,. Edinburgh : Oliver & Boyd 1972.
[9] D. West, *Introduction to graph theory*, Upper Saddle River, NJ : Prentice Hall 1976.
[10] L. Freeman, A. Romney and D. White, (eds.) *Research Methods in Social Network Analysis*, Fairfax, Va: George Mason University Press, 1989.
[11] J. Scott, *Social Network Analysis: A Handbook*, London : Sage, 1991.
[12] S. Wasserman, K. Faust, *Social Network Analysis: Methods and Applications*, Cambridge: Cambridge University Press, 1994.
[13] R. Poulin, M. Boily and B. Masse, Dynamical Systems to Define Centrality in Social Networks. *Social Networks* **22** (3) pp. 187–220, 2000.
[14] M. Granovetter, The Strength of Weak Ties, *The American Journal of Sociology*, **78** (6) May, pp. 1360-1380, 1973.
[15] M. Newman, The Structure and Function of Complex Networks, *SIAM Review*, **45**(2):167-256, 2003.
[16] S. Borgatti, Centrality and Network Flow, *Social Networks* (**27**) pp. 55–71 2005.
[17] R. Albert and A. Barabasi, *Rev. Mod. Phys*. **74**, 47, 2002.
[18] S. Dorogovtsev and J. Mendes, *Evolution of Networks: From Biological Nets to the Internet and WWW*, Oxford University Press, 2005.
[19] S. Macgill, *An appraisal of Q-Analysis*, Working paper 345, School of Geography, University of Leeds, 1982.
[20] J. Johnson, Some Structures and notation of Q-Analysis, *Environment and Planning B*, **8**, pg. 73-86, 1981.
[21] J. Johnson, *The Mathematics of Complex Systems: The Mathematical Revolution Inspired by Computing*, J. Johnson and M. Looms (eds), The Institute of Mathematics and its Applications, Oxford University Press, 1991.
[22] J. Beaumont, A Description of Structural Change in a Central Place System: A Speculation Using Q-analysis, *International Journal of Man-Machine Studies* **20** (6), pp. 567-594, 1984.
[23] L. Duckstein, S. Nobe, Q-Analysis for Modeling and Decision Making, *European Journal of Operational Research* **103**, pp. 411-425, 1997.
[24] T. Fruchterman, E. Reingold, Graph Drawing by Force-directed Placement. *Software, Practice and Experience*, November **21** (1 1), pp. 1129-1164, 1991.
[25] W. de Nooy, A. Mrvar and V. Batagelj, *Exploratory Social Network Analysis with Pajek*, Structural Analysis in the Social Sciences 27, Cambridge University Press, 2005.
[26] W. Christaller, *Central Places in Southern Germany*, Englewood Cliffs, N.J : Prentice-Hall, 1966.

[27] P. Holland, S. Leinhardt, A Method for Detecting Structure in Sociometric Data. *American Journal of Sociology*, **70**, 492-513, 1970.

[28] S. Macgill, Cluster Analysis and Q-analysis, *International Journal of Man-Machine Studies* **20** (6), pp. 595-604, 1984.

[29] M. Newman, Analysis of Weighted Networks, *Phys. Rev. E* **70**, 056131, 2004.

[30] J. Casti, *Connectivity, Complexity and Catastrophe in Large-Scale Systems*. Wiley, New York, 1979.

[31] C. Chin, L. Duckstein, M. Wymore, Factory automation project selection using multicriterion Q-Analysis. *Applied Mathematics and Computation* **46** (2), pp. 107-126, 1991.

[32] R. Milo, S. Shen-Orr, S. Itzkovitz, N. Kashtan, D. Chklovskii, U. Alon. Network Motifs: Simple Building Blocks of Complex Networks. *Science*, **298**, 824-827, 2002.